\documentclass[epj-spec]{svjour}
\usepackage{psfig}

\newcommand{\zabs}{$z_{\rm abs}\,$}

\newcommand{\kms}{km~s$^{-1}\,$}
\newcommand{\ms}{m~s$^{-1}\,$}

\newcommand{\daa}{$\Delta\alpha/\alpha\,$}

\def\la{\;
\raise0.3ex\hbox{$<$\kern-0.75em\raise-1.1ex\hbox{$\sim$}}\; }
\def\ga{\;
\raise0.3ex\hbox{$>$\kern-0.75em\raise-1.1ex\hbox{$\sim$}}\; }

\begin{document}
\title{Bounds on  the fine structure constant
variability from  Fe\,{\sc ii} absorption lines in  QSO spectra}
\author{Paolo Molaro\inst{1}\fnmsep\thanks{\email{molaro@oats.inaf.it}}  
\and Dieter Reimers\inst{2}  
\and Irina I. Agafonova\inst{3} 
\and Sergei A.  Levshakov \inst{3}}
\institute{Osservatorio Astronomico di Trieste, Via G. B. Tiepolo 11,
34131 Trieste, Italy \and Hamburger Sternwarte, Universit\"at Hamburg,
Gojenbergsweg 112, D-21029 Hamburg, Germany 
\and Ioffe Physico-Technical Institute, St. Petersburg, Russia}
\abstract{The Single Ion Differential $\alpha$ Measurement (SIDAM) 
method for measuring  \daa\ 
and its figures of merit are illustrated together with 
the results produced by means of Fe\,{\sc ii} absorption lines 
of QSO intervening systems. The method provides   
\daa~= $-0.12\pm1.79$ ppm (parts-per-million)  at \zabs~= 1.15 
towards HE~0515--4414 and \daa~= $5.66\pm2.67$ ppm at \zabs~= 1.84
towards Q~1101--264, which are so far  the most accurate measurements 
for single systems. SIDAM analysis for 3 systems from the Chand et al. (2004) 
sample provides inconsistent results which we interpret as due to   
calibration errors of the Chand et al. data at the level $\approx$~10 ppm.   
In one system evidence for photo-ionization  Doppler  
shift between Mg\,{\sc ii} and Fe\,{\sc ii} lines  is found. 
This evidence  has important bearings on the Many Multiplet  
method where  the signal for \daa variability 
is carried mainly by systems involving Mg\,{\sc ii} absorbers. 
Some correlations are also found in the Murphy et al. sample which suggest
larger errors than previously reported. 
Thus, we consider unlikely that both the Chand et al.  
and Murphy et al. datasets could provide
an estimate of \daa\  with an accuracy at the level of 1 ppm. 
A new spectrograph like the ESPRESSO project 
will be crucial to make progress in  the astronomical determination of \daa.
} 
\maketitle
\section{Introduction}
\label{intro}
A hypothetical variability of the fine structure
constant $\alpha$ has been  widely discussed in the literature in the last 
few years for its great implications on fundamental physics  
(Dent,  these proceedings) and cosmology (Avelino et al. 2006; Martins 2007, Fujii 2007).

Laboratory experiments are setting  limits on any
changes in $\alpha$ at low energy  to 15 significant figures  
$|\dot{\alpha}/\alpha| =(-2.6 \pm 3.9)\times10^{-16}$ yr$^{-1}$ (Peik et al. 2006) 
and an even lower limit is set by the
fission product analysis of a natural reactor in 
Oklo at  $1.2\times10^{-17}$ yr$^{-1}$ (Gould et al. 2006) occurred at  
redshift $z \simeq 0.2$.
Being linearly extrapolated to higher redshifts ($z > 1$, $\Delta t \sim
10^{10}$ yr), the laboratory bound  leads to 
$|\Delta \alpha/\alpha| \equiv |(\alpha_z - \alpha)/\alpha| <  5 $ ppm
(ppm stands for parts per million, $10^{-6}$) and the Oklo  limit leads to 
$|\Delta \alpha/\alpha|  < 0.1$ ppm. 
However,  the behavior of the  \daa 
variation can be more complex and several  theories
predict very different patterns ranging from slow-rolling to
 oscillating (e.g., Marciano 1984; Mota \& Barrow 2004; Fujii 2005) 
and a linear extrapolation may not be valid on a cosmic time scale.

Variations of $\alpha$ at early cosmological epochs can be probed
at very high redshift $(z \sim 10^{10})$ through the dependence
of primordial light element abundances on fundamental parameters
(e.g., Dent et al. 2007, Coc et al 2007) and at lower redshifts $(z \la 6)$ 
through the measurements of the relative
radial velocity shifts between different metal absorption lines observed
in quasar  spectra. 
Different transitions in molecules and atoms have different sensitivities
to variations of the fundamental physical constants 
(Varshalovich \& Levshakov 1993;  Dzuba et al. 1999). 
This property leads to the Many Multiplet (MM) method  
(Webb et al. 1999; Murphy et al. 2006, hereafter MWF06). 
From the averaging over 143 absorption
systems with redshifts $0.2 < z < 4.2$ identified in the
Keck/HIRES spectra of more than one hundred of QSOs 
Murphy et al. (2004)  obtained \daa~= $-5.7\pm1.1$ ppm.
Chand et al. (2004, hereafter CSPA) 
applying the  MM method to  23 absorption systems from  VLT/UVES spectra   
obtained \daa~= $-0.6\pm0.6$ ppm.  
Observations of two individual systems by means of a slightly different 
methodology  based only on Fe\,{\sc ii} lines, 
which will be described in some detail here,   does not support variability of $\alpha$ as well  (Levshakov 2004; Quast  et al. 2004;  
Levshakov et al. 2005, 2006, 2007a, hereafter L07a).
Thus, currently the case for variability is somewhat controversial 
with Murphy and collaborators claiming negative \daa\  in the past 
at the level of $\approx -6$ ppm  at $5\sigma$ 
confidence level, and  other
groups insisting on non-variability of $\alpha$ at the ppm level. 
 
\begin{table}[t]
\centering
\caption{Atomic data and sensitivity coefficients ${\cal Q}$
for Fe\,{\sc ii} and Mg\,{\sc ii} lines}
\label{tbl-1}
\begin{tabular}{c r@{.}l r@{.}l r@{.}l r@{.}l c}
\hline
\hline
\noalign{\smallskip}
Line & \multicolumn{2}{c}{$\lambda^a_{\rm vac}$, \AA } &
\multicolumn{2}{c}{$f^b$} & 
\multicolumn{2}{c}{${\cal Q}^c_{\rm old}$} & 
\multicolumn{2}{c}{\hspace{-0.2cm}${\cal Q}^d_{\rm new}$}  &
$\Delta {\cal Q}/{\cal Q}$ (\%)  \\
\noalign{\smallskip}
\hline
\noalign{\smallskip}
Fe\,{\sc ii} & 2600&1722 & 0&23878 & 0&035 & 0&0367 & 4$^d$ \\[-2pt]
Fe\,{\sc ii} & 2586&6494 & 0&06918 & 0&039 & 0&0398 & 3$^d$ \\[-2pt]
Fe\,{\sc ii} & 2382&7641 & 0&320   & 0&035 & 0&0369 & 4$^d$ \\[-2pt]
Fe\,{\sc ii} & 2374&4601 & 0&0313  & 0&038 & 0&0394 & 4$^d$ \\[-2pt]
Fe\,{\sc ii} & 2344&2128 & 0&114   & 0&028 & 0&0361 & 26$^d$ \\[-2pt]
Fe\,{\sc ii} & 1611&20034& 0&00138 & 0&018 & 0&0251 & 32$^d$ \\[-2pt]
Fe\,{\sc ii} & 1608&45069& 0&0580  & $-0$&021 & $-0$&0166 & 29$^d$ \\
Mg\,{\sc ii} & 2803&5315& 0&3054  & 0&0034 &\multicolumn{2}{c}{ } 
& 8$^a$ \\[-2pt]
Mg\,{\sc ii} & 2796&3543& 0&6123  & 0&0059 &\multicolumn{2}{c}{ } 
& 5$^a$ \\
\noalign{\smallskip}
\hline
\noalign{\smallskip}
\multicolumn{9}{l}{$^a${\footnotesize
Based on MWF03, Aldenius et al. (2006), L07a.}}\\[-2pt]
\multicolumn{9}{l}{\footnotesize 
$^b$Oscillator strengths $f$ are taken from Morton (2003).}\\[-2pt] 
\multicolumn{9}{l}{\footnotesize 
$^c$Dzuba et al. 1999, 2002. $^d$Porsev et al. 2007.}\\[-2pt]
\end{tabular}
\end{table}

\section{The SIDAM method}

\subsection{Why only Fe\,{\sc ii}?}

The spectroscopic measurability of \daa is based on the fact that the
energy of each line transition depends individually on a change in
$\alpha$. The relative change of the frequency $\omega_0$
due to varying $\alpha$ is proportional to the so-called 
sensitivity coefficient ${\cal Q} = q/\omega_0$. 
The $q$-factors for the resonance UV transitions  were calculated by
Dzuba et al. (2002).  
The updated values corrected for valence-valence and core-valence
correlations and for the Breit interaction were recently reported by
Porsev et al. (2007). 
Both the old and new sensitivity coefficients 
are listed in Table~\ref{tbl-1}.  

The value of \daa\ itself depends on a proper interpretation of
measured relative radial velocity shifts, $\Delta v$, between
lines with different sensitivity coefficients.
It can be shown 
(Levshakov et al. 2006, hereafter L06) 
that in linear approximation ($|\Delta\alpha/\alpha|\ll1$)
\begin{equation}
\frac{\Delta\alpha}{\alpha} = \frac{(v_2 - v_1)}
{2\,c\,({\cal Q}_1 - {\cal Q}_2)} = 
\frac{\Delta v}{2\,c\,\Delta {\cal Q}}\ .
\label{E1}
\end{equation}
 
\begin{figure}[t]
\vspace{0.0cm}
\hspace{0.5cm}\psfig{figure=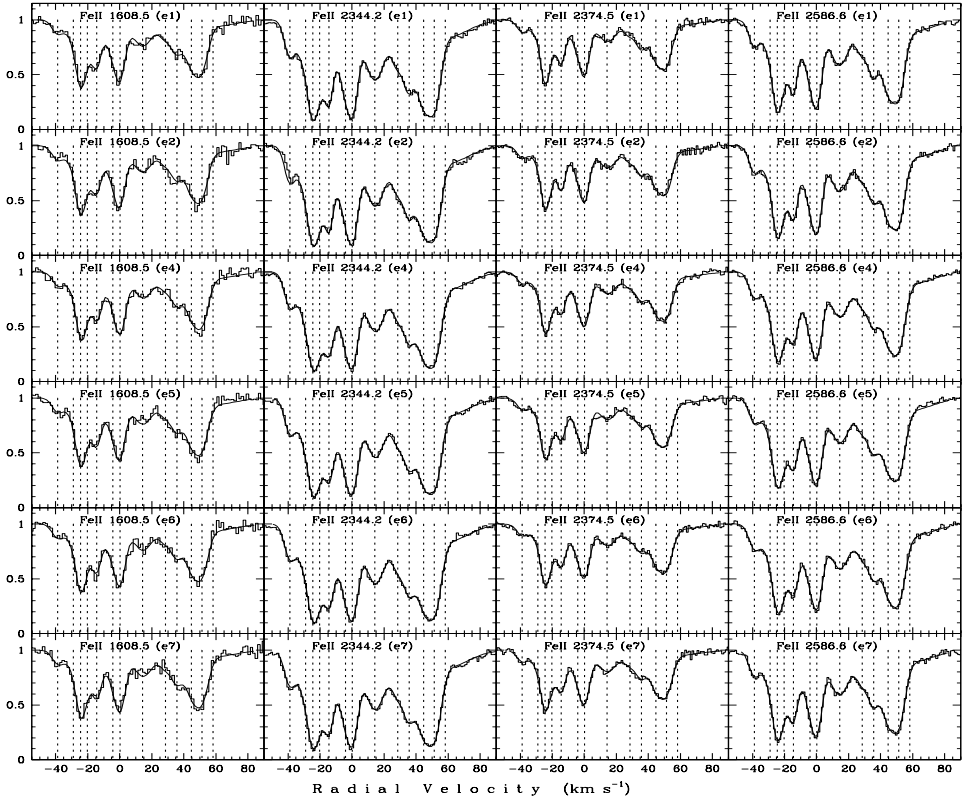,height=11.0cm,width=13.5cm}
\vspace{-0.0cm}
\caption[]{Individual exposures 
of the Fe\,{\sc ii}  lines  of HE 0515--4414. 
Normalized intensities are histograms. 
The over-potted synthetic profiles (smooth curves) are 
calculated from the joint analysis of all Fe\,{\sc ii} profiles. 
The dashed vertical lines mark positions of the 
sub-components. The zero radial velocity is fixed at $z = 1.150965$.
The minimization procedure gives $\chi^2_{\rm min} = 0.9$ per degree of
freedom ($\nu = 2642$).
}
\label{bh1}
\end{figure}

To improve the accuracy of  \daa  measurements  we 
proposed a procedure referred to as 
the Single Ion Differential $\alpha$ Measurement (SIDAM)  
(Levshakov 2004; Levshakov et al. 2005), where the ion is Fe\,{\sc ii}. 
This approach  offers some advantages  with comparison of the MM method, 
namely:

\medskip\noindent
{\bf 1)}\ It is more effective since the Fe\,{\sc ii} 
lines provide positive and negative shifts with 
a larger  $\Delta {\cal Q}$.   Comparing  $\lambda1608$ with  $\lambda2382$ 
or $\lambda2600$ lines,
we have $\Delta {\cal Q} = -0.053$, which  is almost two times 
larger than the same quantity when one compares  
Mg\,{\sc ii} $\lambda\lambda2796, 2803$ and 
Fe\,{\sc ii} $\lambda\lambda2344, 2374, 2382, 2586, 2600$ transitions. 

\noindent
{\bf 2)}\ It is free from
systematics inherent to inhomogeneous ionization. The MM method uses lines
of different ions, which  have different ionization curves. The 
fractional ionization of a given ion depends  on the ionization parameter
$U = n_{\rm ph}/n_{\rm H}$, where $ n_{\rm ph}$ and $n_{\rm H}$ are, 
respectively, the density of the ionizing photons and the gas density. 
Hence, in a cloud with
gas density fluctuations different ions  are Doppler shifted with respect
to each other (Levshakov 2004; Hao et al. 2007). 

\noindent
{\bf 3)}\ It is less sensitive to unknown isotopic abundances. 
The isotope shift between $^{26,24}$Mg\,{\sc ii}  transitions
$3s \rightarrow 3p_{1/2}, 3p_{3/2}$ is $\Delta v_{24-26} \simeq 850$ \ms. 
If the isotope abundance ratio indeed varies with $z$, 
the isotope shifts may imitate the non-zero $\Delta\alpha/\alpha$ value
(Levshakov 1994; Ashenfelter et al. 2004; Kozlov et al. 2004). 
At metallicities of $Z \sim (0.1-1)\,Z_\odot$,~--- typical for the QSO systems
with low ions,~--- the isotope abundances
may not differ considerably from terrestrial.  
Unfortunately,  we do not know the isotope abundances at different redshifts.
The influence of unknown isotopic ratio can be considerably
diminished if only  Fe\,{\sc ii} is used.
The isotopic effect for Fe\,{\sc ii}   
with seven valence electrons in the configurations
$3d^64p$ and $3d^54s4p$ is less pronounced than that
for Mg\,{\sc ii}  with only one valence electron in the configuration $3s$,  
because ($i$) iron is heavier and its isotope structure is more compact, 
and ($ii$) the relative abundance of the leading isotope $^{56}$Fe is 
higher than that of  $^{24}$Mg. 
In fact the terrestrial isotope ratios for Fe are:
$^{54}$Fe: $^{56}$Fe: $^{57}$Fe: $^{58}$Fe = 5.8 : 91.8 : 2.1 : 0.3,  
while those for Mg are 
$^{24}$Mg: $^{25}$Mg: $^{26}$Mg = 79 : 10 : 11.

\noindent
{\bf 4)}\ Less model dependent. All  Fe\,{\sc ii} transitions used in SIDAM have
    identical velocity structure what helps to distinguish the influence
    of hidden blends on the line position measurements. 
    
 \medskip
We have applied the SIDAM methodology only to the brightest QSO 
which provide a better control of instrumental and data reduction systematics. 
When possible we analyze single observations, which allow us 
to follow all individual steps of spectral recording.
For instance calibration exposures should be  taken immediately after 
scientific exposures to minimize the influence of
changing ambient weather conditions which may cause
different velocity offsets in the lamp and QSO spectra if they were
not obtained closely in time.
Variations in temperature can induce color effects in the spectra
since they act differently on the different cross disperses. 
To control possible systematics in radial velocities, the thermal and
pressure stabilities need to be monitored.  
The estimations of Kaufer et al. (2004) for UVES are of 50 \ms\ for 
$\Delta T = 0.3$~K and $\Delta P = 1$ mbar.

\begin{figure}[t]
\vspace{0.0cm}
\hspace{+2.0cm}\psfig{figure=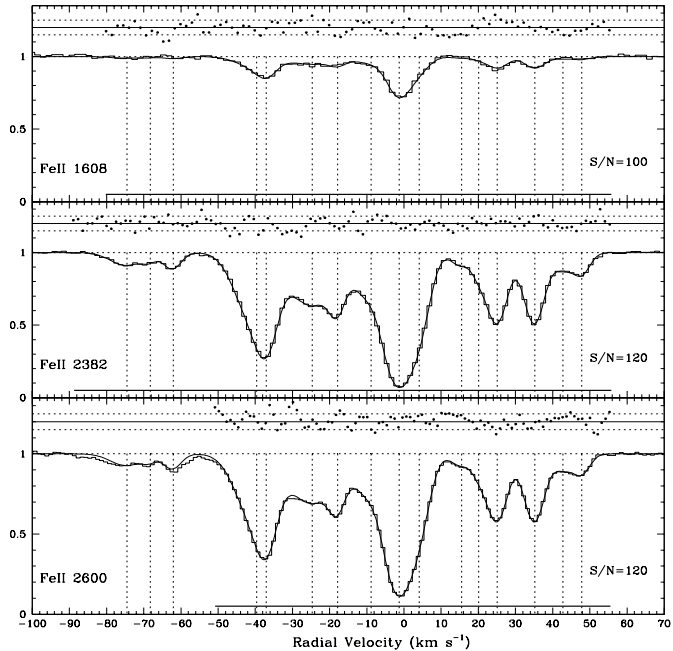,height=9.0cm,width=9.5cm}
\vspace{-0.3cm}
\caption[]{Combined absorption-line spectra of Fe\,{\sc ii}
associated with the \zabs = 1.84 damped Ly$\alpha$ system towards
Q 1101--264 (normalized intensities are shown by histograms).
The zero radial velocity is arbitrarily fixed at $z = 1.838911$.
The synthetic profiles are over-plotted by the smooth curves 
and the dotted vertical lines mark positions of the 
sub-components.
The normalized residuals,
$({\cal F}^{cal}_i-{\cal F}^{obs}_i)/\sigma_i$,
are shown by dots  with   the $1\sigma$ band by  dotted lines. 
Bold horizontal lines mark pixels included in the
optimization procedure. 
The ranges at $v < -50$ \kms and 
at $v \simeq -30$ \kms in the Fe\,{\sc ii} $\lambda2600$ profile
are blended with weak telluric lines. 
The normalized $\chi^2_\nu = 0.901\, (\nu = 257)$. 
}
\label{bh2}
\end{figure}

Using SIDAM, we have almost reached the accuracy of \daa\ 
of $\sim 10^{-6}$ for one absorption system, i.e.
as high as the accuracy of the mentioned above values 
formally obtained from averaging over many absorption systems
where individual \daa values were measured with a considerably
larger than 1 ppm error (see, e.g., Table~\ref{tbl-3} below).

\subsection{Results}
\subsubsection{\daa\ at \zabs = 1.15 towards HE 0515--4414}

The first application of the SIDAM 
method yielded \daa = $-0.4\pm1.9$ ppm at \zabs = 1.15 
towards the $B=15.0$ bright  quasar HE~0515--4414  (Quast et al. 2004). 
This  accuracy was achieved due to unique favorable conditions
such as the brightness of the 
quasar, high spectral resolution ($FWHM \simeq 5$ \kms), 
and the strength of all Fe\,{\sc ii} lines including $\lambda1608$~\AA.

The system  consists of two sub-systems with 
$z_1 = 1.151$  and $z_2 = 1.149$ . 
The former reveals stronger Fe\,{\sc ii}
lines and, hence, provides a higher accuracy of \daa.  
The 6 different spectra of this subsystem analyzed by L06 are reproduced 
in Fig.~\ref{bh1}.
L06 improved the accuracy of \daa\  obtaining  
\daa = $-0.07\pm0.84$ ppm, but 
this analysis did not take into account correlations 
between individual \daa\ values (Levshakov et al. 2007b, hereafter L07b).
If we consider
$k$ Fe\,{\sc ii} pairs with the radial velocity differences
$a_1 = v_2 - v_1, a_2 = v_3 - v_1, \ldots, a_k = v_{k+1} - v_1$,
then the individual values $a_i$ and $a_j$ become correlated with
the correlation coefficient $\rho_{i,j}$ given by
\begin{equation}
\rho_{i,j} = \frac{1}{\sqrt{(1+s^2_i)((1+s^2_j)}}\ ,
\label{EQ8}
\end{equation}
where
$s_i = \sigma_{v_i}/\sigma_{v_1}$, 
$s_j = \sigma_{v_j}/\sigma_{v_1}$, and $v_1$ stands for the 
radial 
velocity of the $\lambda1608$ line.
The errors of the line position measurements
are almost equal for all lines. Thus the ratios in (\ref{EQ8}) are 
$s_i \simeq s_j \simeq 1$, i.e. 
in this case the correlation coefficient  
$\rho_{i,j} \simeq 0.5$.
The  error of the \daa\ estimate from a pair of
Fe\,{\sc ii} lines   is equal to 
$\sigma_{\Delta\alpha/\alpha} \simeq  3.57$ ppm, 
with the new ${\cal Q}$ values reported in Table~\ref{tbl-1}.
The calibration error is 
$\simeq 4$ ppm (L06) that gives the total
$\sigma_{\Delta\alpha/\alpha} = 5.36$ ppm.
Considering that there are 18 pairs  
the error of the mean \daa\ is 1.79 ppm (L07b)
which is 2 times larger than the previous estimate.
Thus, the revised  value of \daa\ at \zabs = 1.15 is
\daa~= $-0.12\pm1.79$ ppm. This determination  
still remains the most accurate individual measurement obtained so far.

A consistent restriction on \daa in this system was obtained
independently by Chand et al. (2006), 
\daa = $0.5\pm2.4$ ppm, who used 
the high resolution pressure and temperature stabilized spectrograph
HARPS mounted on the ESO 3.6~m telescope at the La Silla observatory. 
However, the calibration errors are not considered in the Chand et al.  
total error budget.

\begin{figure}[t]
\vspace{0.0cm}
\hspace{+2.7cm}\psfig{figure=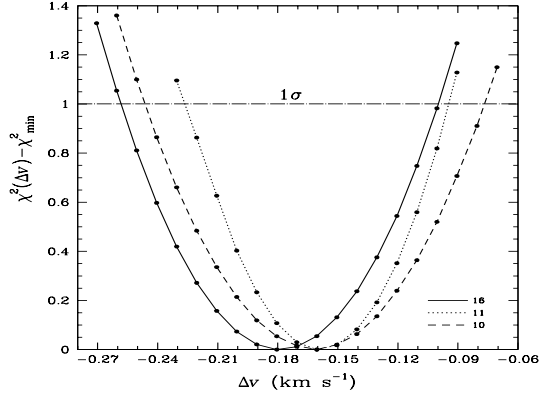,height=6.0cm,width=8.0cm}
\vspace{-0.3cm}
\caption[]{The \zabs = 1.84 systems towards Q~1101--264:
$\chi^2$ as a function of the velocity difference $\Delta v$
between the Fe\,{\sc ii} $\lambda1608$ and $\lambda\lambda2382, 2600$
lines for the 16-, 11-, and 10-component models. 
The corresponding $\chi^2_{\rm min}$ values are equal to 231.631, 135.314, 
and 166.397 (the number of degrees of freedom
    $\nu 257, 188$, and 191, respectively). The minima of the curves give the most probable values of \daa\ :
5.4 ppm (the 16-component model)
and 4.8 ppm (the 11- and 10-component models).
The $1\sigma$ C.L.  gives $\sigma_{\Delta v} = 0.080$ \kms,
0.065 \kms, and 0.085 \kms
or $\sigma_{\Delta \alpha/\alpha} = 2.4$ ppm, 1.9 ppm, and  2.5 ppm for
the 16-, 11-, and 10-component models, respectively
}
\label{bh3}
\end{figure}

\subsubsection{\daa\ at \zabs = 1.84  towards  Q 1101--264} 

The results on the system at \zabs = 1.84  towards  Q~1101--264 
are given in detail in L07a. 
The observations of Q 1101--264 were recorded with the 
VLT UV-Visual Echelle Spectrograph (UVES) on 3 nights in February 2006.    
The resulting normalized and co-added spectra from 9 exposures  are
shown in Fig.~\ref{bh2}.
A signal-to-noise ratio per pixel of S/N~$\ge 100$  
was achieved in the final spectrum. 
Particular care has been adopted in the observations.
The 0.5 arcsec wide slit  provided a  
spectral resolution of  $FWHM \simeq 3.8$ \kms was oriented along 
the parallactic angle. 
Calibration exposures were taken immediately after 
scientific exposures allowing to minimize temperature and pressure changes.  
The pressure variation $\Delta P$ between the QSO exposure and 
the calibration lamp  reached 0.6 mbar in one case, in all other
cases  did not exceed 0.3 mbar, while 
variations of the ambient temperature were always less than 0.1~K. 
The  pixel sizes are of  25 m\AA,
30 m\AA\ and 27 m\AA\ at the positions of $\lambda1608$,
$\lambda2382$ and $\lambda2600$, respectively.
The  wavelength calibration was obtained from  a
ThAr hollow cathode lamp and the Los Alamos table of ThAr lines
(Palmer \& Engleman 1983; de Cuyper \& Hensberge 1998) 
and the residuals of the calibrations  were as small as about 20 \ms.

\begin{figure}[t]
\vspace{0.0cm}
\hspace{+1.5cm}\psfig{figure=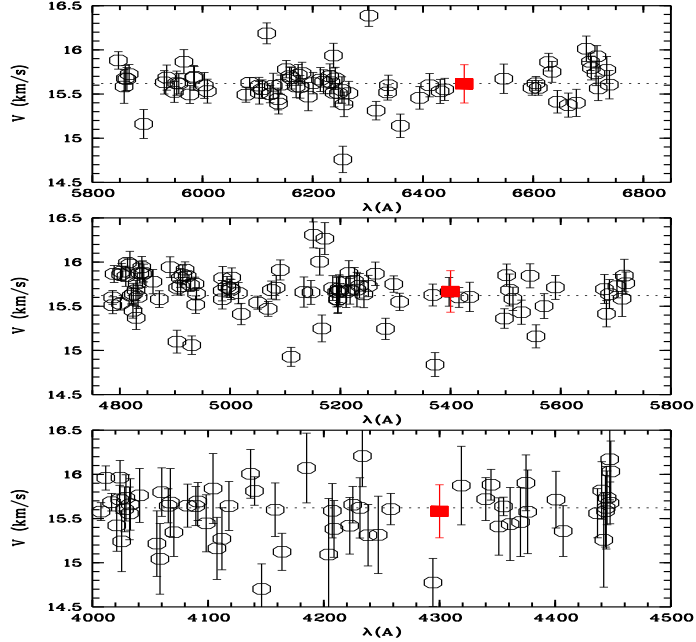,height=9.0cm,width=12cm}
\vspace{0.0cm}
\caption[]{Line shifts of Iris with reference of the solar line wavelengths 
(open circles with $1\sigma$ error bars).
The dotted line shows the expected velocity of the asteroid. 
The panels refer to the three UVES CCDs. Mean values and their dispersion 
(solid squares with error bars)
in the middle of each panel do not show 
any systematic shifts among them.
}
\label{bh4}
\end{figure}

The Fe\,{\sc ii} lines selected for the analysis are those located close 
to the central regions of the corresponding
echelle orders to minimize  possible distortions of the line profiles
caused by the decreasing spectral
sensitivity at the edges of echelle orders. 
We found a good fit of Fe\,{\sc ii} profiles 
for a 16-component model shown in  Fig.~\ref{bh2}
by a  smooth curve and the positions of the sub-components  
marked by dotted lines. The normalized $\chi^2$ per degree of freedom equals
$\chi^2_\nu = 0.901$ ($\nu = 257$).
Since the ${\cal Q}$ values for $\lambda2382$
and $\lambda2600$ are equal, their relative velocity shift 
characterizes the goodness of our wavelength calibration.
We measured a  velocity shift of 20 \ms which is 
comparable with the uncertainty range estimated from the ThAr lines. So,  
we calculate the $\Delta v$ between the  velocity of these two lines 
and that of the  $\lambda1608$ line.
To find the most probable value of $\Delta v$, we fit the absorption lines
with a fixed $\Delta v$, changing $\Delta v$ in the interval from
$-270$ \ms to $-90$ \ms in steps of 10 \ms . For each
$\Delta v$, the strengths of the sub-components, their broadening
parameters and relative velocity positions were allowed to vary 
in order to optimize the fit and thus minimize $\chi^2$.
The $\chi^2$ curve as a function of $\Delta v$ in the vicinity of the
global minimum is shown in Fig.~\ref{bh3}.
The most probable value of $\Delta v$ corresponds to  $-180$ \ms\ 
for the 16-component model. 
The  result is not sensitive to the
number of subcomponents in the model and we have also 
analyzed different models but showing similar radial velocity 
shifts between the $\lambda1608$ and $\lambda2383/2600$ lines.
 The $1\sigma$ errors are equal to 80 \ms, 65 \ms, and 85 \ms\
for the various models examined. 
Adding quadratically the wavelength scale calibration error  of
30 \ms\ between
the blue and red arms,  L07a  obtained
\daa = $5.4\pm2.5$ ppm for the main 16-component model.
With the updated sensitivity coefficients ${\cal Q}_{\rm new}$ the measurement
in L07a  becomes \daa~= $5.66\pm2.67$ ppm. 

This result is quite unexpected and in L07a we have specifically 
investigated those effects which might introduce a 
non-zero difference between the blue and the red lines and thus
simulate a \daa\ variation at the ppm level.
Some systematics are discussed in L07a who were particularly 
concerned of  the possibility that
different velocity offsets occurring  in the blue and red 
frames could cause an  artificial Doppler shift between 
the $\lambda1608$ and $\lambda2382/2600$
lines.  To probe this effect we developed a specific program of   
asteroid observations which are very high precision radial velocity standards. 
The results which are discussed in Molaro et al. (2007) and reproduced 
in Fig.~\ref{bh4} exclude that the shift observed is due
to misalignment of the centering in the two slits of the arms of the 
UVES spectrograph.

\subsubsection{The \zabs = 1.776 system towards  QSO 1331+1704}
  
\begin{figure}[t]
\vspace{0.0cm}
\hspace{3.0cm}\psfig{figure=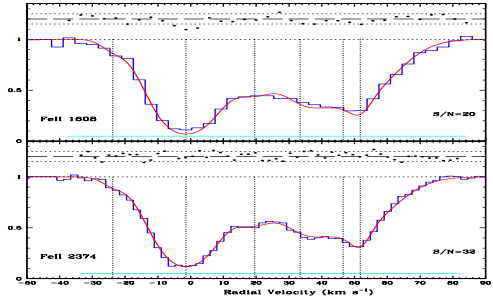,height=4.0cm,width=8.0cm}
\vspace{0.3cm}
\caption[]{The  Fe\,{\sc ii} lines of the  \zabs = 1.776 
system towards QSO 1331+1704. 
}
\label{bh5}
\end{figure}

The \zabs = 1.776 system towards     
QSO 1331+1704 is considered here for the first time for measuring \daa. 
The absorption system shows particularly strong lines with several metal 
lines saturated. This circumstance offers the unique opportunity to apply 
the analysis based on Fe\,{\sc ii} lines to two lines with comparable strength, 
namely the Fe\,{\sc ii} $\lambda1608$ and $\lambda2374$ 
shown in the Fig.~\ref{bh5}.  
The crucial Fe\,{\sc ii} $\lambda1608$ line is normally rather weak and is 
the bottleneck of the  analysis. In this case the line is particularly strong 
making more effective the whole measure. We used two UVES archive images of  
1 hour exposure taken on 18 and 23 Feb 2002.   
The analysis of the archive spectra is shown in Fig.~\ref{bh5}
where the S/N is $\approx$ 20-30.  
We have obtained a velocity difference of 
$\Delta v = -210\pm220$ \ms, which implies \daa~= $5.9\pm6.2$ ppm. 
This result is rather inconclusive due to the large error but obtained 
with very low S/N spectra. A relatively modest increase in the data quality 
will provide easily a much more accurate and significant measure of \daa.  
 
\begin{figure}[t]
\vspace{0.0cm}
\hspace{+3cm}\psfig{figure=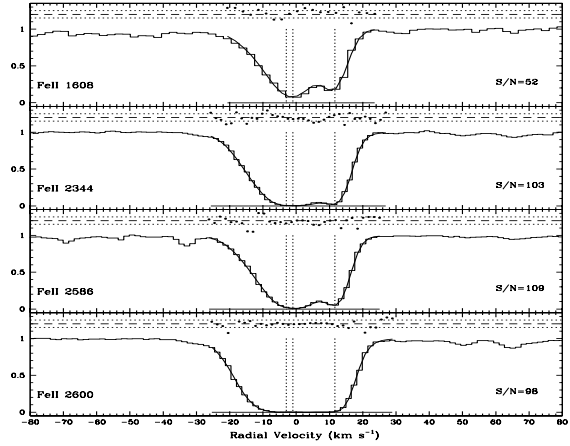,height=6.0cm,width=8.0cm}
\vspace{+0.0cm}
\caption[]{Fe\,{\sc ii} lines of the 
\zabs = 1.43925 system towards HE 1347--2457.
The model profiles are over-plotted
by the smooth curves. The normalized residuals, 
$({\cal F}^{\rm cal}_i - {\cal F}^{\rm obs}_i)/\sigma_i$,  and $1\sigma$ errors
are shown by dots. The vertical lines mark positions of the sub-components. 
Bold horizontal lines mark pixels included in the optimization
procedure. The normalized $\chi^2_\nu = 1.43$ $(\nu = 109)$.
}
\label{bh6}
\end{figure}

\begin{figure}[t]
\vspace{0.0cm}
\hspace{-1.5cm}\psfig{figure=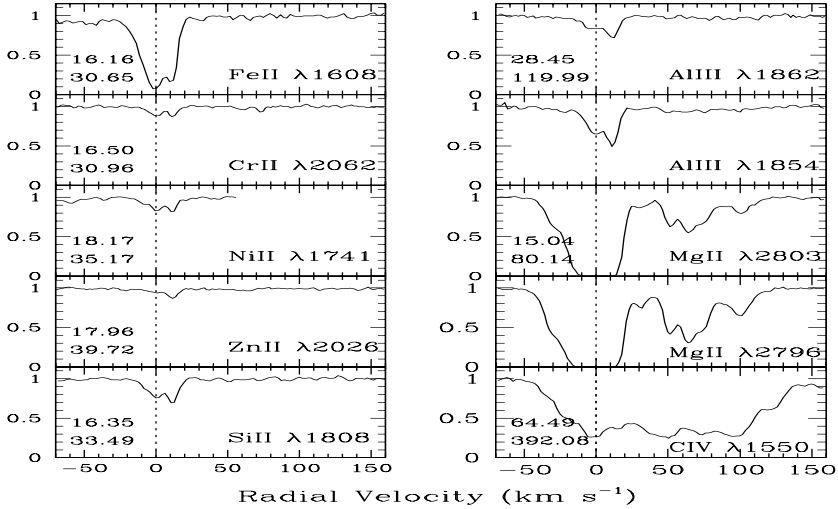,height=12.0cm,width=24.0cm}
\vspace{-5.0cm}
\caption[]{Absorption lines associated with the 
\zabs = 1.439 system towards HE 1347--2457. 
The zero radial velocity is fixed at $z = 1.43925$. 
The ionization potentials (in eV) for the listed ions and their next
ionization state are plotted in each panel. 
Note opposite asymmetry
between subcomponents in  Fe\,{\sc ii}, and 
Cr\,{\sc ii}, Ni\,{\sc ii}, Zn\,{\sc ii}, Si\,{\sc ii}, Al\,{\sc iii} 
in the range $-10 \le v \le 20$ \kms. Besides, Mg\,{\sc ii} traces hotter
gas seen also in Al\,{\sc iii}  and
C\,{\sc iv} in the range $40 \le v \le 130$ \kms.
}
\label{bh7}
\end{figure}

\section{SIDAM analysis of the Chand et al. sample}

From the CSPA sample we selected few systems where the presence of several 
Fe\,{\sc ii}  lines,
including $\lambda1608$,  allowed  us to estimate the value of \daa\
with the SIDAM methodology. The three systems are 
\zabs = 1.439 towards HE 1347--2457, and 
\zabs = 2.185 and \zabs = 2.187 towards HE 0001--2340. 
These systems exhibit also a  very simple line profiles which can be  described
by a minimum number of  components thus   minimizing the uncertainties related 
to model complex line profiles. In addition 
both the systems in HE 0001-2340  reveal unsaturated  Mg\,{\sc ii} 
lines  making it
possible to compare the SIDAM \daa\ value obtained with Fe\,{\sc ii}  lines
with that of the MM method derived from the combination of 
Mg\,{\sc ii} and Fe\,{\sc ii}  lines.
For the  analysis we used the  reduced  spectra courtesy provided by  CSPA, but 
the continuum fitting was made by ourselves by means of the smoothing splines.
We note that the data of CSPA were originally binned by 
$2\times2$ pixels,  providing  corresponding larger pixel sizes.
A detailed analysis of these systems is presented in L07b.

\subsection{HE 1347--2457}
\subsubsection{The \zabs = 1.439 system}

The normalized spectra of four Fe\,{\sc ii}  lines are shown in Fig.~\ref{bh7}.
The spectra have a rather high
signal-to-noise ratio per pixel: S/N~= 52,    
for the Fe\,{\sc ii} $\lambda1608$,  
and $\approx$ 100 for the $\lambda2344$, $\lambda2586$, 
and $\lambda2600$ lines. 
All  Fe\,{\sc ii} lines were fitted simultaneously assuming identical
         velocity structure and allowing for the shift of this structure
         as a whole for every individual  Fe\,{\sc ii} line. The corresponding
         synthetic line profiles for the three component model are
         shown by the smooth curves 
in Fig.~\ref{bh7}.

The sensitivity coefficients of the
{\it red}   Fe\,{\sc ii}  lines $\lambda\lambda2344, 2586, 2600$ are
very close to each other implying that for any \daa\ value these
lines should have the same relative radial velocity.  
Surprisingly, we found that the Fe\,{\sc ii}  $\lambda2600$ line 
shows a significant shift
of $\sim -400$ \ms ($\sim 1/6$ pixel size) as compared to the
Fe\,{\sc ii}  $\lambda\lambda2344, 2586$ lines.  
Considering that this line is close to the border of the 
echelle order where calibration errors may be large,  
we excluded it  from the   calculations.

\begin{figure}[t]
\vspace{0.0cm}
\hspace{+1.0cm}\psfig{figure=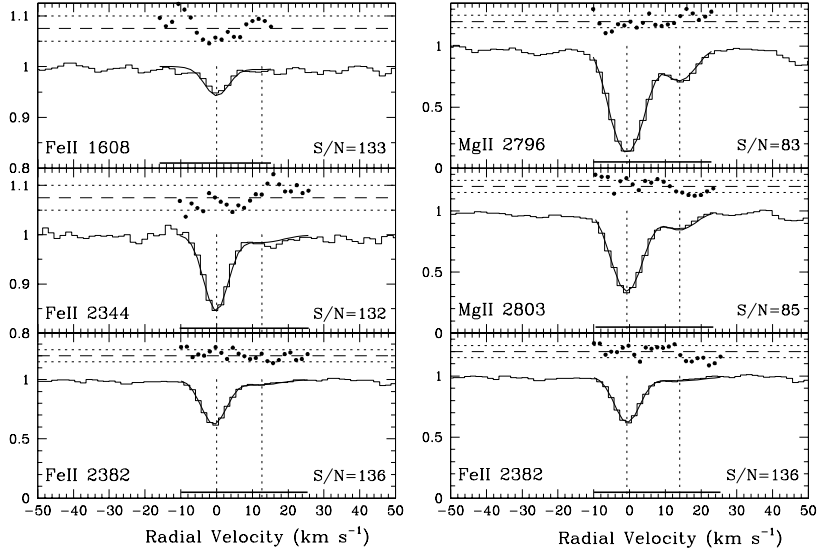,height=12.0cm,width=12cm}
\vspace{-4.0cm}
\caption[]{Fe\,{\sc ii} and Mg\,{\sc i} lines associated with the 
\zabs = 2.185 system towards HE 0001--2340.
The zero radial velocity is fixed at $z = 2.1853$. 
The normalized $\chi^2_\nu$ is equal to: 
0.77 ($\nu = 57$) for the Fe\,{\sc ii}
$\lambda\lambda1608, 2344, 2382$ lines,
and to 1.43 $(\nu = 54)$ for the combined
Mg\,{\sc ii} $\lambda\lambda2796, 2803$ and 
Fe\,{\sc ii} $\lambda2382$ lines.
For other details see capture to Fig.~\ref{bh6}.
}
\label{bh8}
\end{figure}

From the velocity shift between
the Fe\,{\sc ii}  $\lambda1608$ and 
the combined $\lambda\lambda2344, 2586$ lines 
we estimate a  \daa~= $-21.3\pm2.1$ ppm   
(${\cal Q}_{\rm old}$ values from Table~\ref{tbl-1} are used here
and in the following sub-sections in order to compare our calculations
with the CSPA and MWF06 results). 
Taken at face value this result would provide  a detection with a CL 
of 10$\sigma$.  However, both the absolute value of \daa\ and the significance 
level are very unlikely suggesting that  the calibration errors 
are instead responsible for the observed shift.
The accuracy of the above estimate characterizes
solely  the mutual consistency of the lines included
in the $\chi^2$ minimization, but do not consider  the calibration errors.
In  the ESO archive
we could not  find  calibration lamp exposures attached to  the observations of
HE 1347--2457 which means that the
wavelength calibration was performed with
lamps taken at a different moment. 
The large pixel size resulting from binning $2\times2$ 
pixels further deteriorates the accuracy. 
For instance, assuming a  calibration error of $\sim$~1/10 pixel size,  
or uncertainties
of $\sim$~300 \ms and $\sim$~240 \ms for the Fe\,{\sc ii} $\lambda1608$ and 
Fe\,{\sc ii} $\lambda\lambda2344,2586$ lines, respectively,  
we get an error in \daa\ of  $\sigma_{\rm sys} = 11.6$ ppm.  
Considering this as a possible calibration errors and 
adding this error quadratically to the model error, we would obtain
\daa~= $-21.3\pm11.8$ ppm.  
 
We also note that  the Fe\,{\sc ii} and  Si\,{\sc ii}  
lines shown in Fig.~\ref{bh7} have different profiles. 
The simultaneous presence of a strong 
C\,{\sc iv}  and low ionized ions supposes a complex
ionization structure which  may
lead to the velocity shifts between different ions.

\subsection{ HE 0001--2340.}

\subsubsection{The \zabs = 2.185  system}

The normalized spectra of Fe\,{\sc ii} and Mg\,{\sc ii}  lines 
from the \zabs = 2.185 systems towards  HE 0001--2340 are
shown  in Fig.~\ref{bh8}. 
The $\lambda2600$ and $\lambda2586$ lines, not shown,  
are blended with strong telluric absorptions and are not considered.
The signal-to-noise ratio per pixel at the continuum level 
is $\sim 130$ for the Fe\,{\sc ii} lines and $\sim 80$ for 
the Mg\,{\sc ii} lines.
The corresponding pixel sizes (binning $2\times2$ pixels) 
at the positions of the Fe\,{\sc ii}  and Mg\,{\sc ii} lines are of  2.4 \kms.
 
A simple 2-component model is sufficient to describe the 
observed profiles and the corresponding synthetic profiles 
are   shown by the smooth lines in Fig.~\ref{bh8}. 

In this system we use only the $\lambda2382$ line since the $\lambda2344$ 
line is giving a very large velocity shift of 600 \ms\ ($\sim 0.4$ pixel size) 
which we think anomalous (for details, see L07b).
From the analysis of the 
Fe\,{\sc ii}  $\lambda1608$ and Fe\,{\sc ii}  $\lambda2382$  lines we obtain
\daa~= $23.2\pm10.4$ ppm.
As in the preceding system here again
the \daa\ value and its error suggest the presence of  some hidden systematics. 
    
The present system shows unsaturated Mg\,{\sc ii} lines 
with profiles similar to the Fe\,{\sc ii} lines and 
radial velocities of the doublet consistent within $|\Delta v| \le 100$ \ms. 
In the MM method the  Mg\,{\sc ii}  lines are
used as `anchors' due to their low sensitivity to $\alpha$ variations. 
The system under study reveals much more homogeneous ionization structure
as compared to the system 
described above and, hence, allows us to assume small Doppler
shifts between lines of different ions, in particular between 
Fe\,{\sc ii}  and Mg\,{\sc ii}.
The synthetic profiles are shown in
Fig.~\ref{bh8}, right column. The  \daa\  value 
we derive comparing Mg\,{\sc ii} and Fe\,{\sc ii}  is  
\daa~= $3.2\pm2.6$ ppm. However, we note that 
 if an 
error accounting for the calibration uncertainty of the order 
of 11.6 ppm as derived before were added, then
we would obtain \daa~= $3.2\pm18.1$ ppm.

\begin{figure}[t]
\vspace{0.0cm}
\hspace{+1.4cm}\psfig{figure=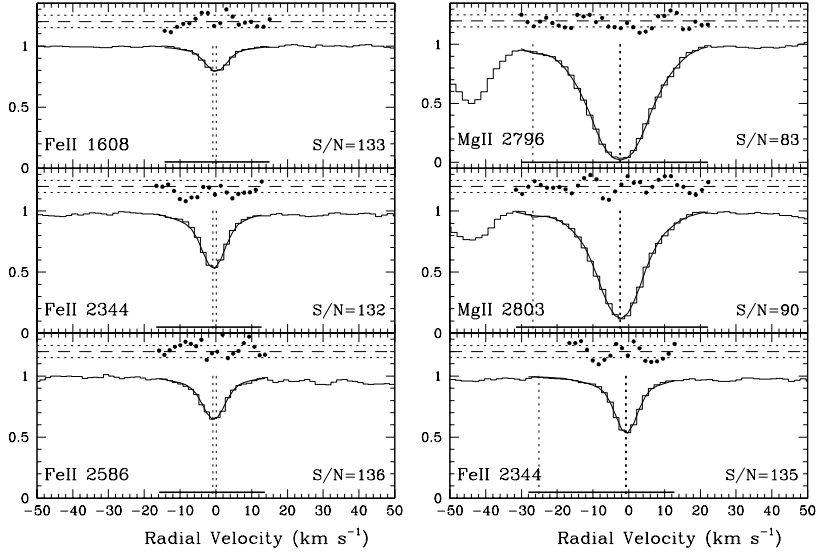,height=12.0cm,width=12cm}
\vspace{-4.0cm}
\caption[]{Fe\,{\sc ii} and Mg\,{\sc ii}
associated with the \zabs = 2.187 system towards HE 0001--2340.
The zero radial velocity is fixed at $z = 2.187155$. 
The normalized $\chi^2_\nu$ is equal to: 
1.52 ($\nu = 50$) for the Fe\,{\sc ii} $\lambda\lambda1608, 2344, 2586$ lines,
and to 1.20 $(\nu = 76)$ for the combined
Mg\,{\sc ii} $\lambda\lambda2796, 2803$ and 
Fe\,{\sc ii} $\lambda2344$ lines.
}
\label{bh9}
\end{figure}

\subsubsection{The \zabs = 2.187 system}

The Fe\,{\sc ii} lines in this system  at  \zabs = 2.187  
shown in Fig.~\ref{bh9} are considerably stronger as
compared to the previous system. 
The relative shift between the red Fe\,{\sc ii} lines and  $\lambda1608$
yields \daa~= $20.8\pm3.1$ ppm. 
At face value  this  result is providing a positive variation of \daa 
with a C.L. of about 7$\sigma$ but of opposite sign of what derived 
towards HE 1347--2457. Again we believe that also this result points 
to the presence of  significant systematic  
errors probably due to the calibration.
Including an estimate of the  calibration uncertainty as before, we would have
\daa~$= 20.8\pm12.0$ ppm.  

The comparison of the position of the Mg\,{\sc ii} and 
Fe\,{\sc ii} lines  reveals something very peculiar.
The fitting of the strongest Fe\,{\sc ii} $\lambda2344$ and 
Mg\,{\sc ii} lines shows that the relative shift between them is extremely large 
and reaches 1.6 \kms. The shift  is of almost 0.7 pixel size and can be even seen 
by eye in Fig.~\ref{bh9}, right column. 
Taking this result at face value,  it would give \daa\  $\sim -90$ ppm. 
This value is even larger than any  conceivable calibration 
error and is suggesting something different.

The system   shows a plenty of lines of
ions in different ionization stages ranging from neutral 
species such as O\,{\sc i} and
C\,{\sc i} to highly ionized such as C\,{\sc iv} and O\,{\sc vi}. 
This suggests that the observed velocity shift between Mg\,{\sc ii} and 
Fe\,{\sc ii} could be caused by the inhomogeneous ionization.

\begin{table*}[t!]
\centering
\caption{Comparison of results from CSPA, MWF06 and this paper
(\daa\ in units of $10^{-6}$)
}
\label{tbl-2}
\begin{tabular}{cccrrcl}
\hline
\hline
\noalign{\smallskip}
QSO & \zabs & \multicolumn{1}{c}{\daa} & \multicolumn{1}{c}{\daa} 
& \multicolumn{1}{c}{\daa} & $\chi^2_\nu$ &\\
 & & \multicolumn{1}{c}{CSPA} 
& \multicolumn{1}{c}{MWF06$^\ast$} & 
\multicolumn{2}{l}{\,\,\,\,\,this work$^{\ast\ast}$}  &  lines used\\
(1) & (2) & (3) & \multicolumn{1}{c}{(4)} & 
\multicolumn{1}{c}{(5)} & \multicolumn{1}{c}{(6)} & 
\multicolumn{1}{l}{(7)} \\
\hline
\noalign{\smallskip}
\noalign{\smallskip}
1347--2457 & 1.493 & & & $-21.3\pm11.8$ & 1.60 & 
{\scriptsize Fe\,{\sc ii}$\lambda\lambda1608, 2344, 2586$. } \\[-2pt]
 & & $0\pm5$ & $-12.7\pm3.6$ & & & 
{\scriptsize Fe\,{\sc ii}$\lambda\lambda2344, 2586$; Si\,{\sc ii}$\lambda\lambda
1526, 1808$. } \\
\noalign{\smallskip}
0001--2340 & 2.185 & & & $23.2\pm15.6$ & 0.71 & 
{\scriptsize Fe\,{\sc ii}$\lambda\lambda1608, 2382$. }\\[-2pt] 
 & & & & $3.2\pm18.1$ & 1.33 & 
{\scriptsize Fe\,{\sc ii}$\lambda2382$;
Mg\,{\sc ii}$\lambda\lambda2796, 2803$. }\\[-2pt] 
 &  & $2\pm3$ & $39.3\pm16.5$ &  & & 
{\scriptsize Fe\,{\sc ii}$\lambda\lambda2344, 2382$;
Mg\,{\sc ii}$\lambda\lambda2796, 2803$; }\\[-2pt] 
 & & & & & & {\scriptsize Si\,{\sc ii}$\lambda1526$; Al\,{\sc ii}$\lambda1670$. }\\ 
\noalign{\smallskip}
 & 2.187 &  &  & $20.8\pm12.0$ & 1.46 &
{\scriptsize Fe\,{\sc ii}$\lambda\lambda1608, 2344, 2586$. }\\[-2pt]
 &  & $-2\pm2$ & $-12.2\pm5.3$ & & &
{\scriptsize Fe\,{\sc ii}$\lambda\lambda2344, 2374, 2586$;
Si\,{\sc ii}$\lambda1526$; }\\[-2pt] 
  & & & & & & 
{\scriptsize  Al\,{\sc ii}$\lambda1670$; Mg\,{\sc ii}$\lambda2803$. }\\
\noalign{\smallskip}
\hline
\noalign{\smallskip}
\multicolumn{7}{l}{$^\ast$ The errors in Col.~4 are
those given in Table~\ref{tbl-1} from MWF06 divided by their 
$\sqrt{\chi^2_\nu}$ values. }\\
\noalign{\smallskip}
\multicolumn{7}{l}{$^{\ast\ast}$ The errors include a  
calibration error of 11.6 ppm, see text. }  
\end{tabular}
\end{table*}

\subsection{Results from the SIDAM analysis of the Chand et al. sample}

The \daa estimated  of CSPA and our  SIDAM measurements 
are summarized in Table~\ref{tbl-2} for comparison. 
The recent reanalysis of MWF06 of the CSPA data, obtained on base of  
Fe\,{\sc ii}, Mg\,{\sc ii},
Si\,{\sc ii}, and Al\,{\sc ii} lines,  are also given in the Table. 

Our analysis with the SIDAM methodology of the CSMP data suggests that 
the VLT/UVES spectra adopted by Chand et al. are affected 
by uncontrolled calibration errors mimicking \daa\ at the level of
10s ppm.  This is best illustrated by  \daa\ values derived from
Fe\,{\sc ii} lines only, but this was already pointed out in L06  
on the basis of Fig.~1(b) of CSPA.
In this figure  of CSPA the 
accuracy of wavelength calibration is checked through the relative velocity
shifts  between the Fe\,{\sc ii} $\lambda2344$ and 
$\lambda2600$ lines. With a good calibration  
the mean $\langle \Delta v \rangle$ should be
consistent with zero since the sensitivity
coefficients of the two Fe lines are very similar. 
On the other hand the figure in the CSPA 
shows  a  significant dispersion of
$\sigma_{\Delta v} \simeq 0.4$ \kms. This scatter   transforms into an  error 
$\sigma_{\Delta\alpha/\alpha} \sim 20$  ppm, 
which is even larger of what assumed here.
Surprisingly, such errors do not show up in the Chand et al. 
results which gives a very small dispersion around zero values.
Recently the same VLT/UVES spectra and the same absorption systems
were re-analyzed by MWF06 who obtained significantly different
values for \daa\ and much larger errors for most of the systems.  CSPA
set a too low tolerance on the halting the calculations while
deriving the $\chi^2(\Delta v)$ curves. As a consequence, the
minima of these curves, i.e. the most probable values of \daa, and their
confidence limits were not accurately estimated. 

The new results for the systems investigated here 
are  given in Table~\ref{tbl-2}.
The results of MWF06 seems to require also some intrinsic scatter or 
larger errors in agreement  with our  analysis  based  only on  
the Fe\,{\sc ii} lines.
Our results  are much in line with those of MWF06 with 
the notable difference of the system at \zabs = 2.187 towards 
HE~0001--2340 where we found  results with opposite sign. 
This can be easily explained with the Doppler shift between 
Mg\,{\sc ii} and Fe\,{\sc ii} ions we detected in this system. 
As noted before this shift is driving \daa towards large negative 
values and this is precisely what found by MWF06.  
The possibility of  such shifts  has important bearings on  
the MM method in general, which, to estimate \daa, 
uses lines of different ions.  Such a possibility has been also 
discussed in some detail in Bahcall et al. (2004).  
Doppler shifts could be of random nature and
would be canceled out if averaged over a large sample of measurements as in 
Murphy et al. (2003, hereafter MWF03). 
However, the uncertainties due to inhomogeneous ionization leading to 
Doppler shifts between different ions
should  be included in the total error budget of the individual
\daa measurements because these errors determine  the weights
and, thus, the values of the weighted mean and its uncertainty.
A priori it is not possible to say whether the averaging
over several  \daa\ values will deliver a true \daa\ value. 
This motivated  a study of the statistical properties of the 
whole \daa\ sample from MWF03 given in L07b and which are briefly 
reported in the following chapter.

\section{ Statistical analysis of the  Murphy et al. sample}

The  accuracy of the \daa\ estimates 
obtained by statistical methods averaging over a large
data array should be checked against systematic effects. 
In the approach by MWF03, the
accuracy of a single \daa\ measurement is of the order of 10 ppm
(see Table~\ref{tbl-3}),  
and only averaging over an hundred of individual
\daa\ values from different redshifts can provide  
an accuracy of the sample mean at 1 ppm level
if the $1/\sqrt{n}$-rule is applicable to the data in question.

\begin{figure}[t]
\vspace{-1.0cm}
\hspace{+1.3cm}\psfig{figure=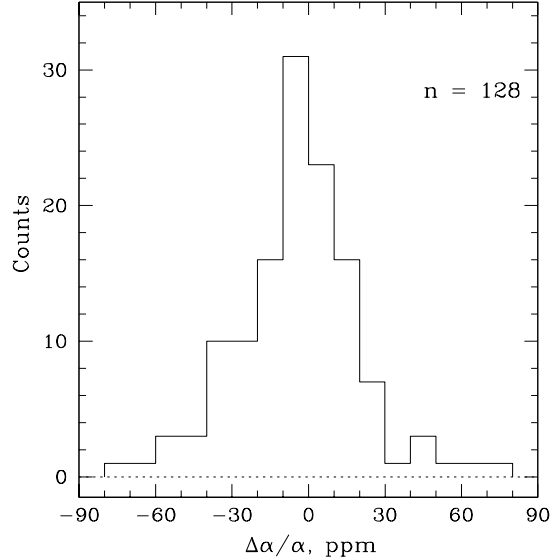,height=12.0cm,width=12.5cm}
\vspace{-2.5cm}
\caption[]{Histogram of the MWF03 sample from Table~\ref{tbl-3}. 
}
\label{bh10}
\end{figure}

The distribution of the MWF03 data 
shown in Fig.~\ref{bh10} reveals pronounced tails. 
In the sample there are also 
some measurements where the negative $\Delta\alpha/\alpha$
     values of several tens of ppm with the significance $\geq 3\sigma$
     are unlikely to be true.
For instance the system at  \zabs = 2.8433 towards QSO 1946+7658 
has a \daa~= $-47.4\pm12.9$ ppm. 
Other notable cases are  the  data points No.~12, 17 and 25 in Table~\ref{tbl-3}.
Since the size of the sample is large enough, 
it would be wise to flag these systems or remove them  
from statistical calculations. The quadratically added
estimate of the unweighted variance is 278.9 ppm$^2$,
whereas the  variance of the individual data points is 552.2 ppm$^2$ 
suggesting that the variance of the MWF03 sample may be underestimated.
Indeed, peculiar clustering is noticeable  
in Table~\ref{tbl-3} where several systems show very 
close \daa\ values (points 37-41, 52, 54, 55, 57, or 78-81). 

A  simple  but nevertheless effective  way to obtain 
a robust estimate is  the $\beta$-trimmed mean where  
$\beta$\% data points are removed from
each side of the distribution  (Hampel et al. 1986). 
Assuming a standard 5\% for $\beta$, 
we obtain a robust 5\%-trimmed mean  of 
$-3.9\pm1.5$ ppm weighted and $-3.7\pm1.7$ ppm unweighted.

Another check is to see whether the data points belonging to 
different subsamples are identically distributed. 
The total sample can be divided into two groups according to  
the data with or without Mg\,{\sc ii}  lines. 
The means of  both subsamples are 
\daa~= $-4.8\pm1.2$ ppm and $-1.1\pm1.7$ ppm more than  $2\sigma$
apart. We note that  all the signal is carried by the first subsample which  
shows negative \daa\ at 
$4\sigma$ confidence level while  
the second one yields  a result  consistent with zero. 

The error of the mean decays with the sample size $n$ as
$\sqrt{n}$   only for independent, i.e. non-correlated, data.  
Any violation of independency leads to the underestimation of the error, which
becomes especially significant for the large size samples.
It should be noted that practically all series of physical measurements 
are somewhat correlated (e.g., Mosteller \& Tukey 1977; Elyasberg 1984;
Hampel et al. 1986; K\"unsch et al. 1993; Hampel 2000).
Since the MWF03 sample is quite large, it is possible to check 
the behavior
of the error of the weighted mean as a function of the sample 
size with the statistical  procedure 
described in Chap.~8 in Hampel et al. (1986).
From the total sample
we form $p$ groups with $n$ elements  randomly selected.  
The sample size of $n_{\rm total} = 128$   allows
$4 \leq p \leq 12$ groups with  $30 \geq n \geq 10$ points in each. For
every group  we calculate  the weighted mean  and then from $p$ values of
the mean we can calculate its variance and standard deviation,
$\sigma_{\rm m}$ (the error of the mean).  
In this way we  obtain the dependence
of $\sigma_{\rm m}$ on the group size. For independent variables this
error goes  as $1/\sqrt{n}$, but if the data
are correlated, then the error decays slower and has a non-zero 
limit at $n \rightarrow \infty$. 

\begin{figure}[t]
\vspace{-2.0cm}
\hspace{3.0cm}\psfig{figure=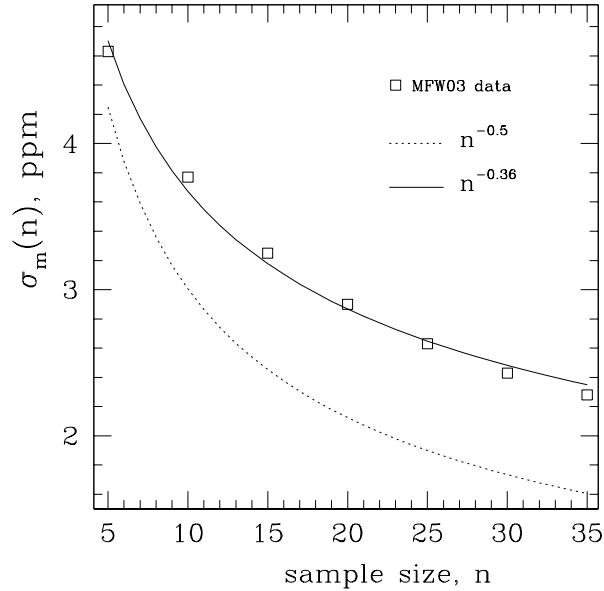,height=11.0cm,width=11.0cm}
\vspace{-0.5cm}
\caption[]{The error of the weighted mean $\sigma_{\rm m}$ 
for the total MWF03 dataset (squares) 
 as a function of the sample size $n$. 
The real error of the mean decays slower than
$n^{-0.5}$ expected for independent data
(see text for detail). 
}
\label{bh11}
\end{figure}

The error $\sigma_{\rm m}$ as a function
of the group size $n$ is shown in Fig.~\ref{bh11} by open squares for the
`raw' MWF03 sample.  
The function $n^{-0.5}$ is plotted by the dashed line.
It is seen that the error of the weighted mean decreases more slowly
than that for the independent data:
in the range $10 < n < 30$ the calculated
points can be approximated by the curve
$n^{-0.425}$  (the smooth line in Fig.~\ref{bh11}). 
This implies that the real error of the mean is at least 1.5 times 
larger than that
estimated on base of the $1/\sqrt{n}$-rule.

\section{ ESPRESSO time}

ESPRESSO stems for Echelle Spectrograph for PREcision Super Stable Observation 
and is a new instrument proposed by a consortium of European institutes for 
the incoherent combined focus of the VLT array (Pasquini, 2008). 
It provides a resolving power of $R = 45000$ collecting the light from 
the 4 VLT units, as an equivalent 16~m telescope,
 or of $R = 180000$ for 1 VLT unit, while incorporating many of the 
innovative solutions of the HARPS spectrograph specialized in hunting 
exoplanets by means of radial velocity variations. 
Thus, ESPRESSO  will provide more photons, more resolution and more 
stability with the promise to improve by at least one order of magnitude 
the accuracy of the \daa\ measurements by means of QSO absorption lines 
(Molaro, 2008). Thus a definitive solution of the case for a variability 
of \daa\ at the level of $\sim 0.1$ ppm will be within reach.

\section{Conclusions}

In this study we have presented the SIDAM methodology  for deriving \daa\ 
by means of the Fe\,{\sc ii} lines observed in the intervening systems in 
QSO spectra.  We have analyzed 3 sets of data used to derive constraints 
on a hypothetical dependence of the fine-structure constant $\alpha$ 
on cosmic time:
($a$) we have revised our previous estimate of \daa\ in the
\zabs = 1.15 system from L06 obtained with the VLT/UVES,
($b$) we have analyzed with SIDAM method 
three absorption systems  included in the sample of CSPA, and 
($c$) considered the statistical properties of 
the sample of MWF03 obtained from Keck/HIRES spectra of QSOs.  
The main results and conclusions are as follows.
\begin{enumerate}
\item
 The estimates of \daa\ are   at
\zabs = 1.15 towards HE 0515--4414 and at \zabs = 1.84
towards Q 1101--264 providing \daa~= $-0.12\pm1.79$ ppm  and
\daa~= $5.66\pm2.67$ ppm, respectively, are the most accurate 
currently estimates of \daa.
A potential systematic  uncertainty which can affect these values is
a relative shift of the wavelength scales in the blue and red arms of 
the UVES where the distant Fe lines are recorded simultaneously.
However, Molaro et al. (2007) by means of  asteroid observations
able to control small shifts at the level of  $\approx$~30 \ms  
revealed no significant radial velocity shifts between the two UVES arms.
\item
To test the CSPA results 
we analyzed  from their sample three absorption systems 
identified in the spectra of
HE 1347--2457 and  HE 0001--2340, where the \daa\ 
can be measured by means of Fe\,{\sc ii} lines only.
The systems
contain the Fe\,{\sc ii} $\lambda\lambda1608, 2344, 2382, 2586$ and 2600
\AA\ lines  which
allows us to estimate \daa\ free from uncertainties related
to the inhomogeneous ionization. The pipeline-reduced
VLT/UVES spectra  used in CSPA show  relative
radial velocity shifts between the Fe\,{\sc ii} lines
mimicking the variability of
$\alpha$ at the level of 10s ppm.
There are no easy ways to evaluate the statistical properties
of these calibration errors and therefore  
the CSPA data are not suitable for the \daa\ estimates at the ppm level.
\item
The statistical analysis of the MWF03 sample shows  
some correlations in the dataset previously overlooked
setting a limit on the achievable accuracy of the weighted mean from 
the applicability of the $\sigma/\sqrt{n}$-rule. 
A robust estimate of \daa\
from a 5\%-trimmed MWF03 sample is \daa~= $-3.9\pm1.5$ ppm
(weighted) or  \daa~= $-3.7\pm1.7$ ppm (unweighted), but it is not clear whether 
the errors of the individual measurements are  still underestimated.
\end{enumerate}

\bigskip\noindent
{\it Acknowledgements.}
This research is supported by
the DFG project RE 353/48-1,
the RFBR grant No.~06-02-16489,
the FASI grant NSh~9879.2006.2.

\begin{table*}[t!]
\centering
\caption{\daa\ with associated $1\sigma$ statistical error
from Keck/HIRES QSO absorption spectra (MWF03).
A symbol `$\ast$' in Cols.~1 and 6 indicates systems with Mg\,{\sc ii} lines. }
\label{tbl-3}
\begin{tabular}{rclrrrclrr}
\hline
\multicolumn{1}{c}{\footnotesize No.} & 
\multicolumn{1}{c}{\footnotesize Object} & \multicolumn{1}{c}{\footnotesize \zabs} & 
\multicolumn{1}{c}{\footnotesize \daa } & 
\multicolumn{1}{c}{\footnotesize $\sigma_{\Delta \alpha/\alpha}$ } & 
\multicolumn{1}{c}{\footnotesize No.} & 
\multicolumn{1}{c}{\footnotesize Object} & \multicolumn{1}{c}{\footnotesize \zabs} & 
\multicolumn{1}{c}{\footnotesize \daa } & 
\multicolumn{1}{c}{\footnotesize $\sigma_{\Delta \alpha/\alpha}$ }  \\[-2pt] 
 & & & \multicolumn{1}{c}{\footnotesize (ppm)} & 
\multicolumn{1}{c}{\footnotesize (ppm)}  & & & & 
\multicolumn{1}{c}{\footnotesize (ppm)} & 
\multicolumn{1}{c}{\footnotesize (ppm)} \\
\hline
{\scriptsize 1}&{\scriptsize $0000-2620$}&{\scriptsize 3.3897}&
{\scriptsize -76.66}&{\scriptsize 32.31}
&{\scriptsize 65}&{\scriptsize $0002+0507$}&{\scriptsize 0.85118}&
{\scriptsize -3.46}&{\scriptsize 12.79}\\[-2pt]
{\scriptsize $\ast$2}&{\scriptsize $1244+3142$}&{\scriptsize 0.85048}&
{\scriptsize -68.97}&{\scriptsize 70.12}
&{\scriptsize $\ast$66}&{\scriptsize $2206-1958$}&{\scriptsize 1.0172}&
{\scriptsize -3.22}&{\scriptsize 7.32}\\[-2pt]
{\scriptsize $\ast$3}&{\scriptsize $1009+2956$}&{\scriptsize 1.1117}&
{\scriptsize -54.61}&{\scriptsize 25.18}
&{\scriptsize $\ast$67}&{\scriptsize $1206+4557$}&{\scriptsize 0.92741}&
{\scriptsize -2.75}&{\scriptsize 7.76}\\[-2pt]
{\scriptsize 4}&{\scriptsize $2230+0232$}&{\scriptsize 1.8585}&
{\scriptsize -54.07}&{\scriptsize 23.40}
&{\scriptsize $\ast$68}&{\scriptsize $0117+2118$}&{\scriptsize 1.0479}&
{\scriptsize -2.23}&{\scriptsize 22.00}\\[-2pt]
{\scriptsize 5}&{\scriptsize $0149+3335$}&{\scriptsize 2.1408}&
{\scriptsize -51.12}&{\scriptsize 29.76}
&{\scriptsize $\ast$69}&{\scriptsize $1206+4557$}&{\scriptsize 0.92741}&
{\scriptsize -2.18}&{\scriptsize 13.89}\\[-2pt]
{\scriptsize 6}&{\scriptsize $1946+7658$}&{\scriptsize 2.8433}&
{\scriptsize -47.43}&{\scriptsize 12.89}
&{\scriptsize 70}&{\scriptsize $1946+7658$}&{\scriptsize 1.7385}&
{\scriptsize -2.12}&{\scriptsize 18.57}\\[-2pt]
{\scriptsize 7}&{\scriptsize $0841+1256$}&{\scriptsize 2.4761}&
{\scriptsize -43.04}&{\scriptsize 28.54}
&{\scriptsize 71}&{\scriptsize $0237-2321$}&{\scriptsize 1.3650}&
{\scriptsize -1.97}&{\scriptsize 5.65}\\[-2pt]
{\scriptsize 8}&{\scriptsize $2233+1310$}&{\scriptsize 3.1513}&
{\scriptsize -40.05}&{\scriptsize 33.01}
&{\scriptsize $\ast$72}&{\scriptsize $0302-2223$}&{\scriptsize 1.0092}&
{\scriptsize -1.89}&{\scriptsize 10.08}\\[-2pt]
{\scriptsize 9}&{\scriptsize $0100+1300$}&{\scriptsize 2.3095}&
{\scriptsize -39.41}&{\scriptsize 24.98}
&{\scriptsize 73}&{\scriptsize $1549+1919$}&{\scriptsize 1.1425}&
{\scriptsize -0.76}&{\scriptsize 6.71}\\[-2pt]
{\scriptsize 10}&{\scriptsize $0201+3634$}&{\scriptsize 2.4563}&
{\scriptsize -37.31}&{\scriptsize 22.85}
&{\scriptsize 74}&{\scriptsize $2359-0216$}&{\scriptsize 2.0951}&
{\scriptsize -0.68}&{\scriptsize 22.11}\\[-2pt]
{\scriptsize $\ast$11}&{\scriptsize $2206-1958$}&{\scriptsize 0.94841}&
{\scriptsize -36.59}&{\scriptsize 18.55}
&{\scriptsize $\ast$75}&{\scriptsize $1248+4007$}&{\scriptsize 0.85452}&
{\scriptsize -0.21}&{\scriptsize 12.68}\\[-2pt]
{\scriptsize $\ast$12}&{\scriptsize $0841+1256$}&{\scriptsize 1.0981}&
{\scriptsize -35.89}&{\scriptsize 12.03}
&{\scriptsize $\ast$76}&{\scriptsize $2344+1228$}&{\scriptsize 1.1161}&
{\scriptsize 0.09}&{\scriptsize 19.63}\\[-2pt]
{\scriptsize $\ast$13}&{\scriptsize $1254+0443$}&{\scriptsize 0.51934}&
{\scriptsize -33.71}&{\scriptsize 32.47}
&{\scriptsize 77}&{\scriptsize $0216+0803$}&{\scriptsize 1.7680}&
{\scriptsize 0.44}&{\scriptsize 12.35}\\[-2pt]
{\scriptsize 14}&{\scriptsize $2231-0015$}&{\scriptsize 2.6532}&
{\scriptsize -33.48}&{\scriptsize 28.27}
&{\scriptsize $\ast$78}&{\scriptsize $1222+2251$}&{\scriptsize 0.66802}&
{\scriptsize 0.67}&{\scriptsize 14.74}\\[-2pt]
{\scriptsize 15}&{\scriptsize $2344+1228$}&{\scriptsize 2.5378}&
{\scriptsize -32.05}&{\scriptsize 20.94}
&{\scriptsize $\ast$79}&{\scriptsize $0117+2118$}&{\scriptsize 0.72913}&
{\scriptsize 0.84}&{\scriptsize 12.97}\\[-2pt]
{\scriptsize $\ast$16}&{\scriptsize $0002+0507$}&{\scriptsize 0.59137}&
{\scriptsize -31.00}&{\scriptsize 24.28}
&{\scriptsize $\ast$80}&{\scriptsize $2145+0643$}&{\scriptsize 0.79026}&
{\scriptsize 0.87}&{\scriptsize 5.89}\\[-2pt]
{\scriptsize $\ast$17}&{\scriptsize $0450-1312$}&{\scriptsize 1.1743}&
{\scriptsize -30.70}&{\scriptsize 10.98}
&{\scriptsize $\ast$81}&{\scriptsize $1421+3305$}&{\scriptsize 0.84324}&
{\scriptsize 0.99}&{\scriptsize 8.47}\\[-2pt]
{\scriptsize 18}&{\scriptsize $1549+1919$}&{\scriptsize 1.8024}&
{\scriptsize -30.50}&{\scriptsize 24.73}
&{\scriptsize $\ast$82}&{\scriptsize $1437+3007$}&{\scriptsize 1.2259}&
{\scriptsize 3.08}&{\scriptsize 14.60}\\[-2pt]
{\scriptsize $\ast$19}&{\scriptsize $1421+3305$}&{\scriptsize 1.1726}&
{\scriptsize -28.44}&{\scriptsize 14.48}
&{\scriptsize $\ast$83}&{\scriptsize $0058+0155$}&{\scriptsize 0.61256}&
{\scriptsize 3.74}&{\scriptsize 11.89}\\[-2pt]
{\scriptsize 20}&{\scriptsize $0347-3819$}&{\scriptsize 3.0247}&
{\scriptsize -27.95}&{\scriptsize 34.29}
&{\scriptsize $\ast$84}&{\scriptsize $0454+0356$}&{\scriptsize 0.85929}&
{\scriptsize 4.05}&{\scriptsize 13.25}\\[-2pt]
{\scriptsize $\ast$21}&{\scriptsize $0055-2659$}&{\scriptsize 1.3192}&
{\scriptsize -26.42}&{\scriptsize 24.57}
&{\scriptsize 85}&{\scriptsize $1425+6039$}&{\scriptsize 2.8268}&
{\scriptsize 4.33}&{\scriptsize 8.27}\\[-2pt]
{\scriptsize $\ast$22}&{\scriptsize $0058+0155$}&{\scriptsize 0.72508}&
{\scriptsize -26.37}&{\scriptsize 35.22}
&{\scriptsize 86}&{\scriptsize $2343+1232$}&{\scriptsize 1.5899}&
{\scriptsize 4.53}&{\scriptsize 11.87}\\[-2pt]
{\scriptsize 23}&{\scriptsize $2231-0015$}&{\scriptsize 2.0653}&
{\scriptsize -26.04}&{\scriptsize 23.23}
&{\scriptsize $\ast$87}&{\scriptsize $0002+0507$}&{\scriptsize 0.85118}&
{\scriptsize 4.94}&{\scriptsize 10.21}\\[-2pt]
{\scriptsize $\ast$24}&{\scriptsize $1634+7037$}&{\scriptsize 0.99010}&
{\scriptsize -21.94}&{\scriptsize 13.43}
&{\scriptsize 88}&{\scriptsize $1223+1753$}&{\scriptsize 2.5577}&
{\scriptsize 5.46}&{\scriptsize 11.99}\\[-2pt]
{\scriptsize $\ast$25}&{\scriptsize $0153+7427$}&{\scriptsize 0.74550}&
{\scriptsize -21.68}&{\scriptsize 7.78}
&{\scriptsize $\ast$89}&{\scriptsize $0841+1256$}&{\scriptsize 1.1314}&
{\scriptsize 5.62}&{\scriptsize 7.87}\\[-2pt]
{\scriptsize 26}&{\scriptsize $0956+1217$}&{\scriptsize 2.3103}&
{\scriptsize -21.61}&{\scriptsize 59.77}
&{\scriptsize 90}&{\scriptsize $0201+3634$}&{\scriptsize 2.4628}&
{\scriptsize 5.72}&{\scriptsize 27.06}\\[-2pt]
{\scriptsize $\ast$27}&{\scriptsize $1107+4847$}&{\scriptsize 1.0158}&
{\scriptsize -20.86}&{\scriptsize 9.34}
&{\scriptsize $\ast$91}&{\scriptsize $1317+2743$}&{\scriptsize 0.66004}&
{\scriptsize 5.90}&{\scriptsize 15.15}\\[-2pt]
{\scriptsize $\ast$28}&{\scriptsize $1107+4847$}&{\scriptsize 0.86182}&
{\scriptsize -20.30}&{\scriptsize 16.32}
&{\scriptsize $\ast$92}&{\scriptsize $0117+2118$}&{\scriptsize 1.3246}&
{\scriptsize 6.95}&{\scriptsize 8.03}\\[-2pt]
{\scriptsize $\ast$29}&{\scriptsize $1626+6433$}&{\scriptsize 0.58596}&
{\scriptsize -19.77}&{\scriptsize 45.29}
&{\scriptsize 93}&{\scriptsize $0201+3634$}&{\scriptsize 2.3240}&
{\scriptsize 7.58}&{\scriptsize 15.92}\\[-2pt]
{\scriptsize $\ast$30}&{\scriptsize $1148+3842$}&{\scriptsize 0.55339}&
{\scriptsize -18.61}&{\scriptsize 17.16}
&{\scriptsize $\ast$94}&{\scriptsize $0528-2505$}&{\scriptsize 0.94398}&
{\scriptsize 7.59}&{\scriptsize 23.35}\\[-2pt]
{\scriptsize 31}&{\scriptsize $1850+4015$}&{\scriptsize 1.9900}&
{\scriptsize -16.63}&{\scriptsize 22.60}
&{\scriptsize 95}&{\scriptsize $0741+4741$}&{\scriptsize 3.0173}&
{\scriptsize 7.94}&{\scriptsize 17.96}\\[-2pt]
{\scriptsize $\ast$32}&{\scriptsize $0450-1312$}&{\scriptsize 1.2294}&
{\scriptsize -14.72}&{\scriptsize 8.36}
&{\scriptsize 96}&{\scriptsize $0528-2505$}&{\scriptsize 2.8114}&
{\scriptsize 8.50}&{\scriptsize 22.55}\\[-2pt]
{\scriptsize $\ast$33}&{\scriptsize $1202-0725$}&{\scriptsize 1.7549}&
{\scriptsize -14.65}&{\scriptsize 21.82}
&{\scriptsize 97}&{\scriptsize $0930+2858$}&{\scriptsize 3.2351}&
{\scriptsize 8.67}&{\scriptsize 17.77}\\[-2pt]
{\scriptsize 34}&{\scriptsize $0757+5218$}&{\scriptsize 2.6021}&
{\scriptsize -13.96}&{\scriptsize 19.55}
&{\scriptsize 98}&{\scriptsize $0019-1522$}&{\scriptsize 3.4388}&
{\scriptsize 9.25}&{\scriptsize 39.58}\\[-2pt]
{\scriptsize $\ast$35}&{\scriptsize $0636+6801$}&{\scriptsize 1.2938}&
{\scriptsize -13.92}&{\scriptsize 6.23}
&{\scriptsize $\ast$99}&{\scriptsize $0450-1312$}&{\scriptsize 1.2324}&
{\scriptsize 10.17}&{\scriptsize 27.52}\\[-2pt]
{\scriptsize $\ast$36}&{\scriptsize $0055-2659$}&{\scriptsize 1.5337}&
{\scriptsize -13.19}&{\scriptsize 10.72}
&{\scriptsize $\ast$100}&{\scriptsize $1634+7037$}&{\scriptsize 0.99010}&
{\scriptsize 10.94}&{\scriptsize 24.59}\\[-2pt]
{\scriptsize $\ast$37}&{\scriptsize $0741+4741$}&{\scriptsize 1.6112}&
{\scriptsize -12.99}&{\scriptsize 17.26}
&{\scriptsize $\ast$101}&{\scriptsize $1107+4847$}&{\scriptsize 0.80757}&
{\scriptsize 11.99}&{\scriptsize 12.22}\\[-2pt]
{\scriptsize $\ast$38}&{\scriptsize $1225+3145$}&{\scriptsize 1.7954}&
{\scriptsize -12.96}&{\scriptsize 10.49}
&{\scriptsize $\ast$102}&{\scriptsize $2231-0015$}&{\scriptsize 1.2128}&
{\scriptsize 12.23}&{\scriptsize 14.65}\\[-2pt]
{\scriptsize $\ast$39}&{\scriptsize $0117+2118$}&{\scriptsize 1.3428}&
{\scriptsize -12.90}&{\scriptsize 9.48}
&{\scriptsize 103}&{\scriptsize $2348-1444$}&{\scriptsize 2.2794}&
{\scriptsize 13.46}&{\scriptsize 41.80}\\[-2pt]
{\scriptsize $\ast$40}&{\scriptsize $1213-0017$}&{\scriptsize 1.5541}&
{\scriptsize -12.68}&{\scriptsize 8.92}
&{\scriptsize 104}&{\scriptsize $1225+3145$}&{\scriptsize 1.7954}&
{\scriptsize 13.52}&{\scriptsize 13.88}\\[-2pt]
{\scriptsize $\ast$41}&{\scriptsize $0000-2620$}&{\scriptsize 1.4342}&
{\scriptsize -12.56}&{\scriptsize 11.67}
&{\scriptsize $\ast$105}&{\scriptsize $2206-1958$}&{\scriptsize 1.0172}&
{\scriptsize 13.54}&{\scriptsize 8.83}\\[-2pt]
{\scriptsize 42}&{\scriptsize $2343+1232$}&{\scriptsize 2.4300}&
{\scriptsize -12.24}&{\scriptsize 21.26}
&{\scriptsize 106}&{\scriptsize $2206-1958$}&{\scriptsize 2.0762}&
{\scriptsize 14.29}&{\scriptsize 30.22}\\[-2pt]
{\scriptsize $\ast$43}&{\scriptsize $0449-1325$}&{\scriptsize 1.2667}&
{\scriptsize -12.12}&{\scriptsize 14.30}
&{\scriptsize 107}&{\scriptsize $0841+1256$}&{\scriptsize 2.3742}&
{\scriptsize 14.35}&{\scriptsize 24.24}\\[-2pt]
{\scriptsize $\ast$44}&{\scriptsize $2343+1232$}&{\scriptsize 0.73117}&
{\scriptsize -12.11}&{\scriptsize 9.75}
&{\scriptsize $\ast$108}&{\scriptsize $1254+0443$}&{\scriptsize 0.93426}&
{\scriptsize 14.85}&{\scriptsize 19.08}\\[-2pt]
{\scriptsize $\ast$45}&{\scriptsize $1421+3305$}&{\scriptsize 0.90301}&
{\scriptsize -9.98}&{\scriptsize 17.83}
&{\scriptsize 109}&{\scriptsize $1223+1753$}&{\scriptsize 2.4653}&
{\scriptsize 16.35}&{\scriptsize 19.19}\\[-2pt]
{\scriptsize 46}&{\scriptsize $2230+0232$}&{\scriptsize 1.8640}&
{\scriptsize -9.98}&{\scriptsize 21.47}
&{\scriptsize $\ast$110}&{\scriptsize $0055-2659$}&{\scriptsize 1.2679}&
{\scriptsize 16.69}&{\scriptsize 27.45}\\[-2pt]
{\scriptsize 47}&{\scriptsize $2343+1232$}&{\scriptsize 2.1711}&
{\scriptsize -9.61}&{\scriptsize 12.95}
&{\scriptsize 111}&{\scriptsize $2231-0015$}&{\scriptsize 2.0653}&
{\scriptsize 17.07}&{\scriptsize 24.35}\\[-2pt]
{\scriptsize $\ast$48}&{\scriptsize $1011+4315$}&{\scriptsize 1.4162}&
{\scriptsize -8.92}&{\scriptsize 5.52}
&{\scriptsize 112}&{\scriptsize $2233+1310$}&{\scriptsize 2.5548}&
{\scriptsize 17.32}&{\scriptsize 63.49}\\[-2pt]
{\scriptsize 49}&{\scriptsize $1442+2931$}&{\scriptsize 2.4389}&
{\scriptsize -8.82}&{\scriptsize 14.73}
&{\scriptsize 113}&{\scriptsize $2206-1958$}&{\scriptsize 1.9204}&
{\scriptsize 18.78}&{\scriptsize 22.05}\\[-2pt]
{\scriptsize 50}&{\scriptsize $0528-2505$}&{\scriptsize 2.1406}&
{\scriptsize -8.53}&{\scriptsize 22.68}
&{\scriptsize 114}&{\scriptsize $0201+3634$}&{\scriptsize 1.9550}&
{\scriptsize 19.89}&{\scriptsize 23.38}\\[-2pt]
{\scriptsize 51}&{\scriptsize $1759+7539$}&{\scriptsize 2.6253}&
{\scriptsize -7.50}&{\scriptsize 25.08}
&{\scriptsize $\ast$115}&{\scriptsize $1248+4007$}&{\scriptsize 0.77292}&
{\scriptsize 21.65}&{\scriptsize 11.91}\\[-2pt]
{\scriptsize $\ast$52}&{\scriptsize $0454+0356$}&{\scriptsize 1.1534}&
{\scriptsize -7.49}&{\scriptsize 17.82}
&{\scriptsize 116}&{\scriptsize $0841+1256$}&{\scriptsize 2.3742}&
{\scriptsize 22.77}&{\scriptsize 43.51}\\[-2pt]
{\scriptsize 53}&{\scriptsize $0207+0503$}&{\scriptsize 3.6663}&
{\scriptsize -7.48}&{\scriptsize 34.68}
&{\scriptsize 117}&{\scriptsize $1244+3142$}&{\scriptsize 2.7504}&
{\scriptsize 24.14}&{\scriptsize 41.10}\\[-2pt]
{\scriptsize $\ast$54}&{\scriptsize $2344+1228$}&{\scriptsize 1.0465}&
{\scriptsize -7.47}&{\scriptsize 15.30}
&{\scriptsize 118}&{\scriptsize $1011+4315$}&{\scriptsize 2.9587}&
{\scriptsize 24.75}&{\scriptsize 26.98}\\[-2pt]
{\scriptsize $\ast$55}&{\scriptsize $1549+1919$}&{\scriptsize 1.3422}&
{\scriptsize -7.40}&{\scriptsize 12.32}
&{\scriptsize $\ast$119}&{\scriptsize $1307+4617$}&{\scriptsize 0.22909}&
{\scriptsize 25.51}&{\scriptsize 53.92}\\[-2pt]
{\scriptsize 56}&{\scriptsize $0241-0146$}&{\scriptsize 2.0994}&
{\scriptsize -7.39}&{\scriptsize 26.75}
&{\scriptsize 120}&{\scriptsize $1055+4611$}&{\scriptsize 3.3172}&
{\scriptsize 27.06}&{\scriptsize 56.77}\\[-2pt]
{\scriptsize $\ast$57}&{\scriptsize $1213-0017$}&{\scriptsize 1.3196}&
{\scriptsize -7.38}&{\scriptsize 7.60}
&{\scriptsize 121}&{\scriptsize $2233+1310$}&{\scriptsize 2.5480}&
{\scriptsize 29.42}&{\scriptsize 52.07}\\[-2pt]
{\scriptsize 58}&{\scriptsize $1626+6433$}&{\scriptsize 2.1102}&
{\scriptsize -7.05}&{\scriptsize 10.68}
&{\scriptsize 122}&{\scriptsize $0757+5218$}&{\scriptsize 2.8677}&
{\scriptsize 38.37}&{\scriptsize 32.88}\\[-2pt]
{\scriptsize 59}&{\scriptsize $1425+6039$}&{\scriptsize 2.7698}&
{\scriptsize -6.88}&{\scriptsize 18.43}
&{\scriptsize 123}&{\scriptsize $1337+1121$}&{\scriptsize 2.7955}&
{\scriptsize 41.03}&{\scriptsize 85.38}\\[-2pt]
{\scriptsize $\ast$60}&{\scriptsize $0201+3634$}&{\scriptsize 1.4761}&
{\scriptsize -6.47}&{\scriptsize 12.19}
&{\scriptsize $\ast$124}&{\scriptsize $0420-0127$}&{\scriptsize 0.63308}&
{\scriptsize 42.11}&{\scriptsize 40.76}\\[-2pt]
{\scriptsize $\ast$61}&{\scriptsize $0841+1256$}&{\scriptsize 1.2189}&
{\scriptsize -5.22}&{\scriptsize 5.42}
&{\scriptsize 125}&{\scriptsize $2359-0216$}&{\scriptsize 2.1539}&
{\scriptsize 43.46}&{\scriptsize 33.38}\\[-2pt]
{\scriptsize 62}&{\scriptsize $1759+7539$}&{\scriptsize 2.6253}&
{\scriptsize -4.92}&{\scriptsize 26.60}
&{\scriptsize 126}&{\scriptsize $1215+3322$}&{\scriptsize 1.9990}&
{\scriptsize 56.48}&{\scriptsize 37.64}\\[-2pt]
{\scriptsize $\ast$63}&{\scriptsize $0940-1050$}&{\scriptsize 1.0598}&
{\scriptsize -4.53}&{\scriptsize 15.72}
&{\scriptsize 127}&{\scriptsize $1132+2243$}&{\scriptsize 2.1053}&
{\scriptsize 63.23}&{\scriptsize 36.22}\\[-2pt]
{\scriptsize $\ast$64}&{\scriptsize $0823-2220$}&{\scriptsize 0.91059}&
{\scriptsize -3.94}&{\scriptsize 6.09}
&{\scriptsize $\ast$128}&{\scriptsize $0119-0437$}&{\scriptsize 0.65741}&
{\scriptsize 71.23}&{\scriptsize 45.99}\\[-2pt]
\noalign{\smallskip}
\hline
\noalign{\smallskip}
\end{tabular}
\end{table*}

\end{document}